\documentclass[aps,amsmath,amssymb,nofootinbib,superscriptaddress,showpacs,
floatfix,twocolumn,prb,10pt]{revtex4-1}

\usepackage{latexsym}
\usepackage{graphicx}
\usepackage{times,psfrag,subfigure}
\usepackage{amsmath}
\usepackage{dcolumn}
\usepackage{bm,bbm}       
\usepackage{color}
\usepackage{latexsym,amsmath,amssymb,bm,euscript}
\usepackage{chngcntr}
\counterwithin*{paragraph}{subsection}

\def\Tr{\mathop{\mathrm{Tr}}}

\def\sgn{\mathop{\textrm{sgn}}}

\newcommand\cln{\color{black}}
\newcommand{\beq}{\begin{equation}}
\newcommand{\eeq}{\end{equation}}
\newcommand{\beqarray}{\begin{eqnarray}}
\newcommand{\eeqarray}{\end{eqnarray}}

\newcommand{\eq}[1]{Eq.~(\ref{#1})}  
\newcommand{\fig}[1]{Fig.~\ref{#1}}  
\newcommand{\figs}[1]{Figs.~\ref{#1}}  
\newcommand{\Tab}[1]{Tab.~\ref{#1}}  
\newcommand{\bsigma}{\mbox{\boldmath$\sigma$}}
\newcommand{\btau}{\mbox{\boldmath$\tau$}} 
\newcommand{\PART}[1]{{\bf{\emph{#1}}}:} 
\newcommand{\physrep}{Phys. Rep.} 
\newcommand{\prx}{Phys. Rev. X}

\graphicspath{{Graphics/}}

\begin{document}

\allowdisplaybreaks

\title{Edge instabilities of topological superconductors}

\date{\today}

\author{Johannes S. Hofmann}
\email{jhofmann@physik.uni-wuerzburg.de} 
\affiliation{Institut f\"ur Theoretische Physik und Astrophysik,
Universit\"at W\"urzburg, Am Hubland, D-97074 W\"urzburg, Germany}
\affiliation{Max-Planck-Institut f\"ur Festk\"orperforschung,
  Heisenbergstrasse 1, D-70569 Stuttgart, Germany} 

\author{Fakher F. Assaad}
\affiliation{Institut f\"ur Theoretische Physik und Astrophysik,
Universit\"at W\"urzburg, Am Hubland, D-97074 W\"urzburg, Germany}

\author{Andreas P. Schnyder}
\email{a.schnyder@fkf.mpg.de}
\affiliation{Max-Planck-Institut f\"ur Festk\"orperforschung,
  Heisenbergstrasse 1, D-70569 Stuttgart, Germany}

\begin{abstract}
Nodal topological superconductors display zero-energy Majorana flat bands at generic edges. The flatness of these edge bands, which is protected by time-reversal and translation symmetry, gives rise to an extensive ground-state degeneracy.
Therefore, even arbitrarily weak interactions lead to an instability of the flat-band edge states towards time-reversal and translation-symmetry-broken phases, which lift the ground-state degeneracy. 
We examine the instabilities of the flat-band edge states of $d_{xy}$-wave superconductors by performing a mean-field analysis in the Majorana basis of the edge states. The leading instabilities are Majorana mass terms, which correspond to coherent superpositions of particle-particle and particle-hole channels in the fermionic language.
We find that attractive interactions induce three different mass terms. One is a coherent superposition of imaginary $s$-wave pairing and current order, and another combines a charge-density-wave and finite-momentum singlet pairing.
Repulsive interactions, on the other hand, lead to ferromagnetism together with spin-triplet pairing at the edge.
Our quantum Monte Carlo simulations confirm these findings and demonstrate that these instabilities occur even in the presence of strong quantum fluctuations.
We discuss the implications of our results for experiments on cuprate high-temperature superconductors.
\end{abstract}

\date{\today}

\pacs{02.70.Ss, 03.65.vf, 71.27.+a, 73.20.-r, 74.20.Rp, 74.50.+r}


\maketitle

\PART{Introduction}
The discovery of topological insulators\cite{Hasan2010,Qi2011} has led to the insight  that 
nontrivial band topologies  
can give rise to exotic surface states\cite{Hasan2010,Qi2011,chiu_review15}. 
Particularly interesting are topological flat-band surface states, since their large
density of states enhances correlation effects\cite{Matsumoto1995, Matsumoto1995a, Fogelstroem1997, Timm2015, Honerkamp2000, Potter2014, KopninVolovikPRB11, graphene_edge_magnetism, Feldner2010, Feldner2011, Roy2014, tangFu_natPhys14}.
Surface states with a (nearly) flat dispersion can occur both 
in topological semimetals\cite{tangFu_natPhys14,graphene,ChiuSchnyder14} and in nodal topological superconductors (SCs)\cite{Tanaka2012,Matsuura2013,schnyder_review15,Schnyder2011}. However, only in the latter systems is 
the flatness of the surface states protected by symmetry\cite{RyuHatsugaiPRL02,Tanaka2009,Schnyder2011}.
That is, time-reversal symmetry (TRS), particle-hole symmetry (PHS),
and translation symmetry ensure that the surface states are pinned at zero energy, resulting in a band of neutral Majorana 
fermions. 

These Majorana bands exist in one- or two-dimensional regions of the surface Brillouin zone, 
which are bounded by the projections of the superconducting nodes. 
Hence, the number of zero-energy surface states grows linearly or quadratically with the length of the system,
leading to a diverging density of states at zero energy and an extensive ground-state degeneracy. Since this is in
violation with the third law of thermodynamics,
 even arbitrarily weak interactions cause a singular perturbation of the Majorana flat bands, giving rise to  novel symmetry-broken states at the surface\cite{KopninVolovikPRB11, Honerkamp2000, Potter2014, graphene_edge_magnetism, tangFu_natPhys14, Feldner2010, Feldner2011, Roy2014, Li2013}.
Due to the flat-band character and the low dimensionality of the boundary, these
symmetry-broken states are subject to strong fluctuations. Therefore, it is necessary
to use methods beyond mean-field (MF) theory\cite{kauppila_arXiv_15} 
in order to analyze the surface instabilities. 

In this Rapid Communication, we employ a mean-field analysis together with continuous-time quantum Monte Carlo (QMC) simulations\cite{Rubtsov2005,Gull2011,Pavarini2014}
 to examine the interaction effects on the Majorana flat-band edge states of  
$d_{xy}$-wave superconductors. These edge states
are experimentally realized in cuprate
 high-temperature superconductors\cite{Lee2006,Scalapino1995}
and have been observed in tunnel junction experiments on normal-metal YBa$_2$Cu$_3$O$_{7-x}$ junctions.
At intermediate temperatures, these measurements show a sharp zero-bias peak\cite{Geerk1988,Lesueur1992,Covington1996,kashiwaya_PRB_95,alff_PRB_97,wei_PRL_98,kashiwaya_tanaka_rep_prog_phys_00} 
that arises due to the diverging density of states of the edge states.
Upon further cooling, the observed zero-bias peak splits into two\cite{Covington1997,Krupke1999}, which 
is interpreted as a sign of spontaneous TRS breaking\cite{footnoteA}.
This was examined by several MF studies\cite{Matsumoto1995, Matsumoto1995a, Fogelstroem1997, Potter2014, Honerkamp2000, Timm2015},
which found that for attractive interactions the order parameter
develops imaginary $s$-wave components near the boundary, while
for repulsive interactions edge ferromagnetism (FM) is induced. 

The purpose of this Rapid Communication, is to go beyond these previous MF calculations and to conduct a systematic examination of all possible instabilities of the flat-band edge states using
(i) a mean-field analysis in the \emph{Majorana basis} of the edge states and (ii)
continuous-time QMC simulations which take into account fluctuation effects.
Interestingly, we find that for repulsive interactions, the  FM instability is coherently mixed
with a spin-triplet pairing instability. For attractive interactions, on the other hand, the  $s$-wave pairing instability
is combined with current order and similarly charge-density-wave (CDW) instability, whose wavevector $Q$ corresponds to nesting between the flat bands, is mixed with finite-momentum singlet pairing.
We show that for attractive interactions  
and at half filling 
long-range order is established at the edge at $T=0$.
Our findings are relevant for experiments on 
cuprate high-temperature superconductors and we provide experimental setups to test these unique signatures of Majorana flat bands.

\begin{table*}
\begin{minipage}{0.7\textwidth}
\begin{tabular}{|c|c|c|}
\hline 
non-zero vev & mass term & fermionic correlation along interacting edge \\ 
\hline 
$\left\langle S_0^{x,y\vphantom{(\Psi)}} \right\rangle$ 
& $ \frac{1}{2} \sum_{k_\parallel=0}^\pi \Gamma^\dagger_{k_\parallel}  m^{x,y}_{k_\parallel} \tau^{x,y} \Gamma^{\phantom{\dagger}}_{k_\parallel} $ 
& $\sum_j \left[ a_0 S^{x,y}_j + b_1(\Delta_j^{b;x,y} + {\Delta_j^{b;x,y}}^\dagger) \right] + \cdots$ \\ 
\hline 
$\left\langle S_0^{z\vphantom{(\Psi)}} \right\rangle$ 
& $ \frac{1}{2} \sum_{k_\parallel=0}^\pi \Gamma^\dagger_{k_\parallel}  m^z_{k_\parallel} \tau^z \Gamma^{\phantom{\dagger}}_{k_\parallel} $ 
& $\sum_j \left[ a_0 S^z_j - b_1(\Delta_j^{b,z} + {\Delta_j^{b,z}}^\dagger) \right] + \cdots$  \\ 
\hline 
$\left\langle S_\pi^{x(\Psi)} \right\rangle$ 
& $\frac{1}{2} \sum_{k_\parallel=0}^\pi \tilde{\Gamma}^\dagger_{k_\parallel}  g^x_{k_\parallel} \tau^x \tilde{\Gamma}^{\phantom{\dagger}}_{k_\parallel}$ 
& $\sum_j (-1)^j \left[ a_0 (\Delta_j^{s} + {\Delta_j^{s}}^\dagger)  + b_1 n^{b}_j\right]  + \cdots$ \\ 
\hline 
$\left\langle S_0^{y(\Psi)} \right\rangle$ 
& $\frac{1}{2} \sum_{k_\parallel=0}^\pi \tilde{\Gamma}^\dagger_{k_\parallel}  (-g^y_{k_\parallel}) \tau^z \tilde{\Gamma}^{\phantom{\dagger}}_{k_\parallel}$ 
& $\sum_j  \left[ - i a_0 (\Delta_j^{s} - {\Delta_j^{s}}^\dagger) + b_1 J_j \right] + \cdots$ \\ 
\hline 
$\left\langle S_\pi^{z(\Psi)} \right\rangle$ 
& $\frac{1}{2} \sum_{k_\parallel=0}^\pi \tilde{\Gamma}^\dagger_{k_\parallel}  g^z_{k_\parallel} \tau^y \tilde{\Gamma}^{\phantom{\dagger}}_{k_\parallel}$ 
& $\sum_j (-1)^j \left[ a_0 n_j - b_1(\Delta_j^{b,s} + {\Delta_j^{b,s}}^\dagger) \right] + \cdots$ \\ 
\hline 
\end{tabular} 
\end{minipage}
\begin{minipage}{0.29\textwidth}
\begin{tabular}{|c|c|}
\hline 
operator & definitions \\ 
\hline 
$n_{j}$ & $c^{\dagger}_{j} \sigma^0 c^{\phantom{\dagger}}_{j}$  \\ 
\hline 
${\bf{S}}_j^{}$ & $c^\dagger_{j} \frac{\bsigma}{2} c^{}_{j}$  \\ 
\hline 
$\Delta^s_{j}$ & $-c^{\phantom{\dagger}}_{j;\uparrow} c^{\phantom{\dagger}}_{j;\downarrow}$  \\ 
\hline 
$n^b_{j}$ &  $c^{\dagger}_{j} \frac{\sigma^0}{2} c^{\phantom{\dagger}}_{j+1} + h.c.$  \\ 
\hline 
$J_{j}$ & $c^{\dagger}_{j} \frac{i \sigma^0}{2} c^{\phantom{\dagger}}_{j+1} + h.c. $ \\ 
\hline 
$\Delta^{b,s}_{j}$ & $c_{j}^{\mathrm{T}}i\tau_y\frac{\tau^0}{2} c_{j+1} $
\\ 
\hline 
$\mathbf{\Delta}^{b}_{j}$ & $c_{j}^{\mathrm{T}}i\tau_y\frac{\btau}{2} c_{j+1} $
\\ 
\hline 
\end{tabular} 
\end{minipage}
\caption{Summary of all possible MF channels at half filling. The left table lists possible vacuum expectation values, their associated masses for the edge states,  and the characterizing fermionic correlations. We use $\Gamma_{k_\parallel}^\dagger=(\gamma_{k_\parallel}^\dagger,-i\,s_{k_\parallel}\,\gamma_{-k_\parallel}^{\ })$ and $\tilde{\Gamma}_{k_\parallel}^\dagger = (\gamma_{k_\parallel}^\dagger ,-i\,s_{k_\parallel}\,\gamma_{k_\parallel-\pi}^{\dagger })$ . The $(\cdots)$ indicate additional operators on higher-order bonds. \label{Tab:MFchannel}}
\end{table*}

\PART{Model}
We start from a phenomenological description of
a single-band $d_{xy}$-wave SC given in terms of
the Bogoliubov--de~Gennes Hamiltonian $\mathcal{H}_0 = \sum_{\bf k}\Psi_{\bf    k}^{\dagger}H({\bf k})\Psi^{\ }_{\bf k}$,
with the Nambu spinor $\Psi_{\bf k} = ( c_{{\bf k} \uparrow},  c^\dag_{-{\bf k} \downarrow}  )^{\mathrm{T}}$ and
\begin{eqnarray}
\label{Bulk_model}
H({\bf k}) = \begin{pmatrix}
\varepsilon_{\bf k} &
\Delta_{\bf k} \cr
\Delta^{\ast}_{\bf k} &  
- \varepsilon_{-\bf k} 
\end{pmatrix} .
\end{eqnarray}
Here, $c^{\dag}_{{\bf k}\sigma}$ denotes the electron creation operator with
spin $\sigma$ and momentum ${\bf k}=(k_\parallel=k_x,k_\perp=k_y)^{\mathrm{T}}$, anticipating a later introduced ribbon geometry with open boundary conditions in the $y$ direction.
The normal part of the Hamiltonian describes a two-dimensional square 
lattice with nearest-neighbor hopping $t$ and chemical potential $\mu$, hence  
$\varepsilon_{\bf k}
= -2t\, ( \cos k_\parallel + \cos k_\perp) - \mu$.
The SC order parameter $\Delta_{\bf k}= \Delta_{d_{xy}} \sin k_\parallel \sin k_\perp$ contains only spin-singlet pairing of amplitude $\Delta_{d_{xy}}$.

To discuss the topology of this two-dimensional (2D) nodal system, we interpret $H\left( k_\parallel, k_\perp \right)$ as a set of fully gapped chains $H_{k_\parallel}(k_\perp)$, indexed by $k_\parallel$. Each subsystem falls into class BDI and its topology is classified by a winding number\cite{Schnyder2008,Matsuura2013,Kitaev2001,Schnyder2012}.  The subsystem exhibits a nontrivial bulk topology if $2\left|t\right|>\left|\mu_{k_\parallel}\right|$ and $\Delta_{k_\parallel}\neq0$ and hosts protected zero energy edge states (created by $\gamma^\dagger_{k_\parallel}$) once open boundary conditions for the perpendicular direction $k_\perp$ are imposed. Here we use the shorthand notations $\mu_{k_\parallel}=\mu+2t\cos(k_\parallel)$ and $\Delta_{k_\parallel}=\Delta_{d_{xy}}\sin(k_\parallel)$. The interested reader may find a more detailed discussion of the topology and the protected edge states in Sec.~I of Ref.~\cite{supplement}.

To study the correlation effects among Majorana states, we include a Hubbard interaction along the top edge ($i_{\perp,0}=1$) by refining the Hamiltonian to $\mathcal{H} = \mathcal{H}_0 + \mathcal{H}_{\mathrm{int}}$ with
\beq
\label{Interaction}
\mathcal{H}_{\mathrm{int}} = -\frac{2U}{3L} \sum_{q_\parallel} {\bf S}_{-q_\parallel}{\bf S}_{q_\parallel} =\frac{2U}{3L} \sum_{q_\parallel} {\bf S}^{(\Psi)}_{-q_\parallel}{\bf S}^{(\Psi)}_{q_\parallel} 
\eeq
in terms of the physical spin operator ${\bf{S}}_q^{} = \sum_{k_\parallel} c^\dagger_{k_\parallel} \frac{\bsigma}{2} c^{}_{k_\parallel+q}$ or a pseudospin operator ${\bf{S}}_q^{(\Psi) } = \sum_{k_\parallel} \Psi^\dagger_{k_\parallel} \frac{\btau}{2} \Psi^{}_{k_\parallel+q}$.

Unless stated otherwise, we use $(t, \mu, \Delta_{d_{xy}}, L_\perp ) = (1.0, 0.0, 1.0, 10^2)$.

\begin{figure*}
\centering
\includegraphics[clip,angle=0,width=.95\textwidth]{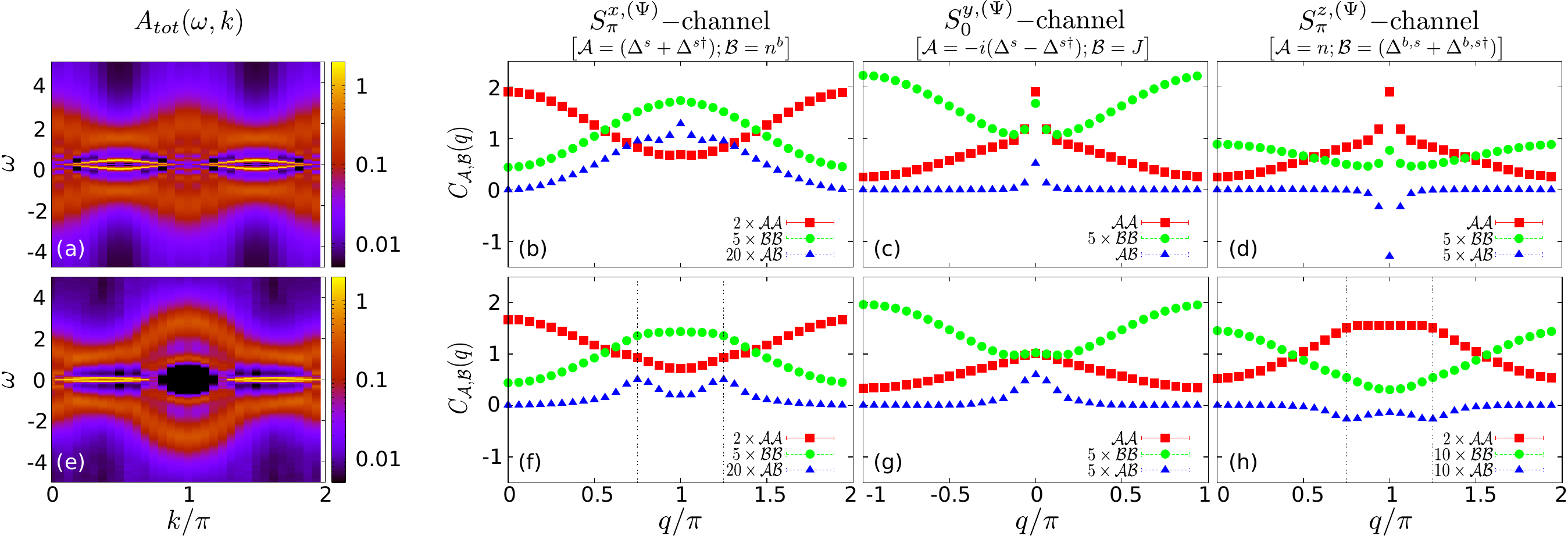}
\caption{\label{PlotAtt}  
(Color online) 
We present the single particle spectrum $A_{\mathrm{tot}}(\omega,k)$ and equal-time correlation functions for attractive interactions with $L=32$ and $(U,\mu,\beta/t)=(-2,0,100)$ in the top [(a)--(d)] and $(U,\mu,\beta/t)=(-1,-0.586,50)$ in the bottom [(e)--(h)].
The edge states have been gapped out and instabilities can be identified in all three ${\bf S}^{(\Psi)}$ channels as defined in \Tab{Tab:MFchannel}.
} 
\end{figure*} 

\PART{Mean-field considerations}
Let us examine some MF decouplings before presenting the numerical simulations.
We restrict our discussion to the interacting edge sites and drop the index $i_\perp = i_{\perp,0}$ for readability. All derivations assume half filling $\mu=0$.

\emph{Repulsive interaction}: 
In the presence of repulsive interactions one expects FM instabilities, hence we approximate $\mathcal{H}_{\mathrm{int}}$
by a MF decoupling ${\bf{mS}}_0^{}$. Projecting on the Majorana states generates the mass term
\beq
\frac{1}{2} \sum_{k_\parallel=0}^\pi  \Gamma^\dagger_{k_\parallel}  {\bf{m}}_{k_\parallel} \btau \Gamma^{\phantom{\dagger}}_{k_\parallel}+\cdots , \label{MajoranaMassRep}
\eeq
with $\Gamma_{k_\parallel}^\dagger=(\gamma_{k_\parallel}^\dagger,-i\,s_{k_\parallel}\,\gamma_{-k_\parallel}^{\ })$, $s_{k_\parallel}=\sgn(t\Delta_{k_\parallel})$, and ${\bf{m}}_{k_\parallel}=\phi^2_{k_\parallel}(i_{\perp,0})\,{\bf{m}}$. The $(\cdots)$ represent edge-bulk and bulk-bulk contributions. This reproduces the edge splitting terms known from Ref.~\cite{Potter2014}.
Due to the $SU(2)$-spin symmetry of the Hamiltonian, the orientation $\bf{m}$ remains arbitrary. A nonzero value $\left|\bf{m}\right|$ breaks time-reversal and spin-rotation symmetry.

To make the connection with the QMC simulations, we express Eq.~\eqref{MajoranaMassRep} in terms of fermionic correlations along the edge (see  \Tab{Tab:MFchannel} derived in Sec.~II  of Ref.~\cite{supplement}).
Due to the chiral structure of the edge states, a non-zero mass $| {\bf m}|$ corresponds to a coherent superposition of FM and spin-triplet SC, where the in-plane (out-of-plane) components are parallel (antiparallel) aligned. In this analysis, we decomposed the $k_\parallel$ dependence of $\phi^4_{k_\parallel}$ in harmonics. Accordingly, there will be further contributions on next-nearest neighbor and higher-order bonds, oscillating  between normal and SC operators.

\emph{Attractive interactions}: As indicated by \eq{Interaction}, the transformation $c_{\bf k}\rightarrow\Psi_{\bf k}$ renders $U>0$ repulsive in terms of ${\bf{S}}_q^{(\Psi) }$. Hence, we expect pseudo-magnetic instabilities. First focusing on homogeneous instabilities ($Q=0$), we find that  ${\bf{S}}_0^{(\Psi) }$ projected on the Majorana states is vanishing except for the $y$ component.
Therefore only a condensation of $S_0^{y,(\Psi) }$ gaps the edge spectrum.
Including inhomogeneous order (i.e., $Q\ne0$) opens additional channels. It is natural to study those wave vectors $Q$ that maximize the nesting between edge states with opposite chiral eigenvalue.
At half filling, this fixes $Q=\pi$.
Projecting ${\bf{S}}_\pi^{(\Psi) }$ on the Majorana states generates nontrivial operators for the $x$ and $z$ but a vanishing $y$ component, complementary to $Q=0$.
 
The MF decoupling ${\bf{g}}(S_\pi^{x,(\Psi) }, S_0^{y,(\Psi) }, S_\pi^{z,(\Psi) })^{\mathrm{T}}$ generates the Majorana masses
\beq
\frac{1}{2} \sum_{k_\parallel=0}^\pi \tilde{\Gamma}^\dagger_{k_\parallel} \left( g^x_{k_\parallel} \tau^x + {\bf{\tilde{g}}}_{k_\parallel} \btau \right) \tilde{\Gamma}^{\phantom{\dagger}}_{k_\parallel}+\cdots , \label{MajoranaMassAttr}
\eeq
with $\tilde{\Gamma}_{k_\parallel}^\dagger=(\gamma_{k_\parallel}^\dagger,-i\,s_{k_\parallel}\,\gamma_{k_\parallel-\pi}^{\dagger })$, $g^x_{k_\parallel}=\phi^2_{k_\parallel}(i_{\perp,0}) g^x$, and ${\bf{\tilde{g}}}=\phi^2_{k_\parallel}(i_{\perp,0}) {\bf{g}} \times {\bf{e_x}} $.
At half filling, we make use of a sublattice symmetry $U^{\mathrm{SL}}= \sum_{k_\parallel,i_\perp}(-1)^{i_\perp} \Psi^\dagger_{k_\parallel,i_\perp} \frac{\tau^x}{2}\Psi^{}_{k_\parallel+\pi,i_\perp}$. This symmetry generates rotations in the $(y,z)$ plane that change the orientation of $ \bf{\tilde{g}} $, but leave $\left| \bf{\tilde{g}} \right|$ and $g_x$ invariant. Hence, there is a competition between these two channels.
Interestingly, the sublattice symmetry combines a time-reversal and a translation-symmetry-breaking sector in $ \bf{\tilde{g}} $.

As before, we rewrite \eq{MajoranaMassAttr} in terms of fermionic operators, the result of which is shown in \Tab{Tab:MFchannel}.
We obtain linear superpositions of normal and SC operators. $S_\pi^{x,(\psi)}$ combines finite-momentum $s$-wave pairing with a bond-density-wave instability, $S_0^{y,(\psi)}$ contains complex $s$-wave SC and edge current operators, and $S_\pi^{z,(\psi)}$ includes a CDW instability and finite-momentum singlet SC on nearest-neighbor bonds.

Doping the system breaks the symmetry $U^{\mathrm{SL}}$. As a result, the constraint on $S^{y,(\psi)}$ and $S^{z,(\psi)}$ is lifted, which allows for a competition between both channels. As the bulk nodes move away from $0$ or $\pi$, the nesting wave vector $Q$ decreases and we expect instabilities in the $S^{x,(\psi)}$ and $S^{z,(\psi)}$ channel at $Q<\pi$. 

\PART{Method}
We use a continuous-time QMC method in the interaction expansion\cite{Rubtsov2005,Gull2011}. To incorporate $d$-wave SC, we formulate the simulation in the Nambu basis. We perform the calculations using an effectively one-dimensional Green's function, which contains the degrees of freedom of the two-dimensional bulk states\cite{Hohenadler2011,Hohenadler2012,Hohenadler2014}. For more details on the QMC method we refer the reader to Sect.~III of Ref.~\cite{supplement}.
The single particle spectra $A_{\mathrm{tot}}(\omega,k)=-(2\pi)^{-1}\sum_\sigma\mathrm{Im}G_\sigma(\omega,k)$ are extracted from the time-ordered Green's function $\langle c^\dagger_{k,\sigma}(\tau) c^{\phantom{\dagger}}_{k,\sigma}(0)  \rangle$ using the stochastic maximum entropy method\cite{Sandvik1998,Beach2004}. 
To identify the mentioned Majorana masses, we determine equal-time correlation functions
\begin{equation}
C_{\mathcal{A},\mathcal{B}}(q) = \frac{1}{L} \sum_{n,n'}^L e^{iq(n-n')} 
\left( \langle \mathcal{A}^{\dagger}_{n}\mathcal{B}^{\phantom{\dagger}}_{n'} \rangle - \langle \mathcal{A}^{\dagger}_{n} \vphantom{\mathcal{B}^{\phantom{\dagger}}_{n'}} \rangle \langle \mathcal{B}^{\phantom{\dagger}}_{n'} \rangle \right) .
\label{CorrFuncts}
\end{equation}

\PART{Results} 
The QMC simulation is sign-problem free for attractive interactions ($U=-2$) at half filling such that we can perform a scaling analysis and extrapolate to the thermodynamic limit. Doping and/or repulsive interaction introduce a sign problem. Hence, we only extract leading instabilities for $L=32$ and $U=\pm 1$. \cln

\emph{Attractive interactions:} We first study the system at half filling and $\beta/t=100$. The single particle spectrum is shown in \fig{PlotAtt}(a).
We observe that the zero-energy flat bands develop a dispersion and gap out.
Hence the interaction along the edge dynamically generates Majorana masses.
The masses discussed above can generate this spectrum and lead to an unique set of coherent fermionic correlations.
Figures~1(b)--1(d) suggest instabilities associated with both $\left|g^x\right|\neq 0$ ($S^{x(\Psi)}_\pi$-channel) and $\left|{\bf{\tilde{g}}}\right|\neq 0$ ($S^{y(\Psi)}_0$- and $S^{z(\Psi)}_\pi$-channel). Each nontrivial cross correlation confirms the expected coherent superposition of normal and SC correlations.
\begin{figure}
\begin{minipage}{0.55\columnwidth}
\vspace{0.2cm}
\centering
\includegraphics[clip,angle=0,width=.9\columnwidth]{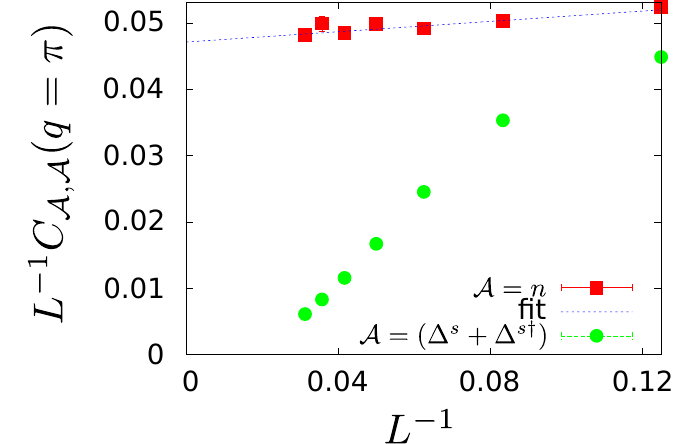}
\end{minipage}
\begin{minipage}{0.43\columnwidth}
\caption{\label{FermionicInstabilities}  
(Color online) 
Finite size scaling of $\left| \tilde{{\bf g}} \right|^2$ (red) and $\left| g^x \right|^2$ (green) with fixed $\beta=\frac{50}{8}L$ in red and green. The extrapolation for $\mathcal{A}=n$ suggests long-range order ($\left| \tilde{{\bf g}} \right| \neq 0$) at $T=0$.
}  
\end{minipage}
\end{figure}
Figure~2 visualizes the scaling behavior of the correlation function for the CDW, representing the ${\bf{\tilde{g}}}$~channel, and for s-wave singlet SC, representing $g^x$-channel. The data suggest long-range order at $T=0$ in the ${\bf{\tilde{g}}}$~channel, whereas $g^x$ vanishes.
Observe that we employed the enhanced symmetry of the zero-energy subspace (i.e., the chiral nature of the edge states) to derive the fermionic correlation functions associated to each Majorana mass. However, this symmetry does not manifest itself for the order parameter as it would unify the three channels by promoting the $U(1)$ sublattice symmetry to a $SU(2)$~symmetry.

Doping the system removes the sublattice symmetry and allows a competition between the $S^{y(\Psi)}_0$- and $S^{z(\Psi)}_Q$-channels.
Figure 1(e) shows the single particle spectrum and
we again observe a splitting of the flat-band.
Once more the correlation function in \figs{PlotAtt}(f)--(h) show instabilities in all channels, which are best seen in the cross correlations between normal and SC contributions.
The doping of $\mu=-0.586$ induces $Q=\pm\frac{3}{4}\pi$, which explains the instabilities in the $S^{x(\Psi)}$ and $S^{z(\Psi)}$~channel.

\emph{Repulsive interactions:}
\begin{figure}
\centering
\includegraphics[clip,angle=0,width=\columnwidth]{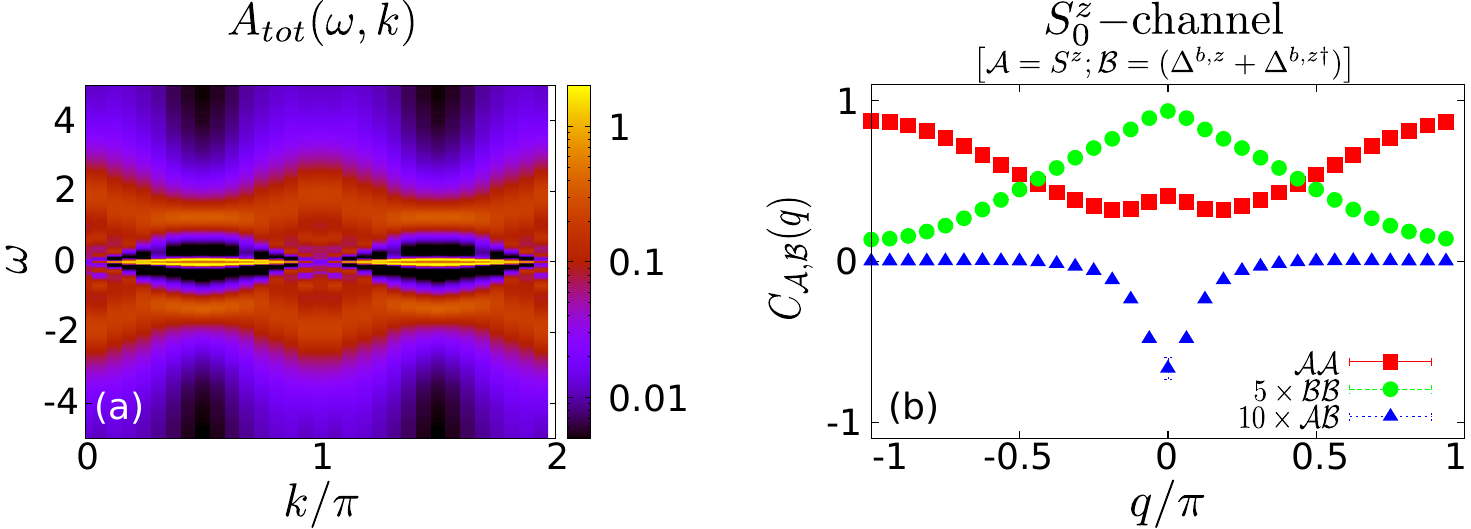}

\caption{\label{repSingleSpec}  
(Color online) 
We present the single particle spectrum $A_{\mathrm{tot}}(\omega,k)$ (a) and the correlation functions (b) for $L=32$ and $\beta/t=100$. The edge states have been gapped out and the FM is coherently mixed with triplet SC.
}  
\end{figure}
The results for $L=32$ and $\beta=100$ are shown in~\fig{repSingleSpec}. Again, the Majorana states are gapped out. We can confirm edge FM as the leading instability\cite{Potter2014}. In contrast to  previous studies, however, we find from the MF analysis that the FM is coherently mixed with a (anti)parallel polarized triplet SC. This is well confirmed by the correlation functions depicted in~\fig{repSingleSpec}(b).

\PART{Discussion} 
Previous MF studies proposed ferromagnetism or additional $is$-wave pairing\cite{Matsumoto1995, Matsumoto1995a, Fogelstroem1997, Potter2014, Honerkamp2000, Timm2015} along the edge as leading instabilities. Our unbiased QMC results, together with a refined MF analysis, show, however, that $is$-wave pairing and the FM are coherently mixed with current order and spin-triplet pairing, respectively. 
That is, the order parameters are linear superpositions of both normal conducting and superconducting operators,
as shown by the nontrivial cross correlations (e.g., between the spin polarization and triplet pairing) in 
Figs.~\ref{PlotAtt}(b)--(d), \ref{PlotAtt}(f)--(h), and \ref{repSingleSpec}(b).
Indeed, the key insight from the MF analysis is that the instabilities correspond to Majorana mass terms, 
which in the fermionic language correspond to superpositions of particle-particle and particle-hole channels. 
This coherent superposition is a direct consequence of the chiral nature of the Majorana edge state. If there were both chiralities at one edge, the linear combination would be lost. Hence probing the coherence between the different fermionic order parameters provides useful information about the character of the edge states.

The agreement of the MF considerations and the QMC analysis is remarkable considering that the former completely neglected all bulk state effects. We effectively projected $\mathcal{H}_{\mathrm{int}} \sim (e^\dagger+b^\dagger)(e+b)(e^\dagger+b^\dagger)(e+b)$ to $e^\dagger ee^\dagger e$ and ignored all bulk state contributions. Here, $b$ and $e$ represent bulk and edge degrees of freedom, respectively, where $e$ has definite chirality. In principle, higher-order contributions could allow for chirality flipping pair-scattering terms which might also split the edge states\cite{Queiroz2014}. The $d_{xy}$-wave SC is nodal and therefore hosts gapless excitations in its bulk. Accordingly, there is no separation in energy which justifies these approximations.

To detect the coherence between the FM and triplet SC in the Majorana masses, relevant for repulsive interactions (the most likely scenario for underdoped YBCO cuprate), we propose Josephson current measurements in SC-FM-SC junctions\cite{Sheyerman2015}. It would be useful to compare the currents in junctions where the interface is aligned along the (110) direction (with edge states) to those in junctions with an interface
along the (100) direction (no edge states).
The polarization direction of the FM can be controlled in this setup by applying an external magnetic field. 
We expect that in this junction the ferromagnetic part of the Majorana mass is  aligned with the FM of the junction. This also fixes the polarization of the triplet component to be either parallel or antiparallel to the FM, depending on the orientation (see \Tab{Tab:MFchannel}). This polarization direction is expected to strongly influence the tunneling probability and therefore the  Josephson current.
By varying the polarization of the FM, one can manipulate the relative phase in the superposition between the FM and the triplet pairing, such that we would not only detect the presence of additional triplet pairing along the edge but also infer information about the coherence between the different components.

In the presence of attractive interactions, the CDW order will be pinned by impurities or by the underlying lattice\cite{Lee1974}. Thereby, charge modulations in STM  should be observable. 

\PART{Summary}
In this Rapid Communication, we have studied instabilities of chiral flat-band Majorana fermions in topological SCs using QMC. We have confirmed the FM instability for repulsive interactions beyond the mean-field level. Our analysis points out that any normal conducting order is coherently mixed with a SC counterpart due to the Majorana nature of the edge states, for example FM and triplet SC. This mixing should open up possibilities to detect the instabilities experimentally. In the case of attractive interactions, the system exhibits long-range order at half filling and $T=0$, namely, CDW combined with finite-momentum extended $s$-wave pairing and complex $s$-wave SC in superposition with current order. In a doped system, these two orders compete with each other and the numerical data suggest an instability towards SC mixed with spontaneous edge currents.

\begin{acknowledgments}
The authors thank P.~Brouwer, P.~Brydon, F.~Goth, M.~Hohenadler, E.~Khalaf, R.~Queiroz, C.~Timm, and M.~Weber for useful discussions. J.--H. and F.--A. are supported by the German Research Foundation (DFG), under DFG-SFB~1170 ``ToCoTronics" (Project C01) and DFG-FOR~1162 (AS120/6-2).
We thank the J\"ulich Supercomputing Centre for generous allocation of CPU time.
\end{acknowledgments}


%


\clearpage

\appendix

\begin{center}
\textbf{
\large{Supplemental Material for}}
\vspace{0.4cm} 

\textbf{
\large{
``Edge instabilities of topological superconductors" } 
}
\end{center}

\vspace{0.1cm}

\begin{center}
\textbf{Authors:}  
Johannes S. Hofmann,
Fakher F. Assaad,
and
Andreas P.\ Schnyder
\end{center}

\section*{I.~~~Topology, Edge States and Mass Terms}
\label{appendixSurStates}

\begin{figure}
\centering
\includegraphics[clip,angle=0,width=\columnwidth]{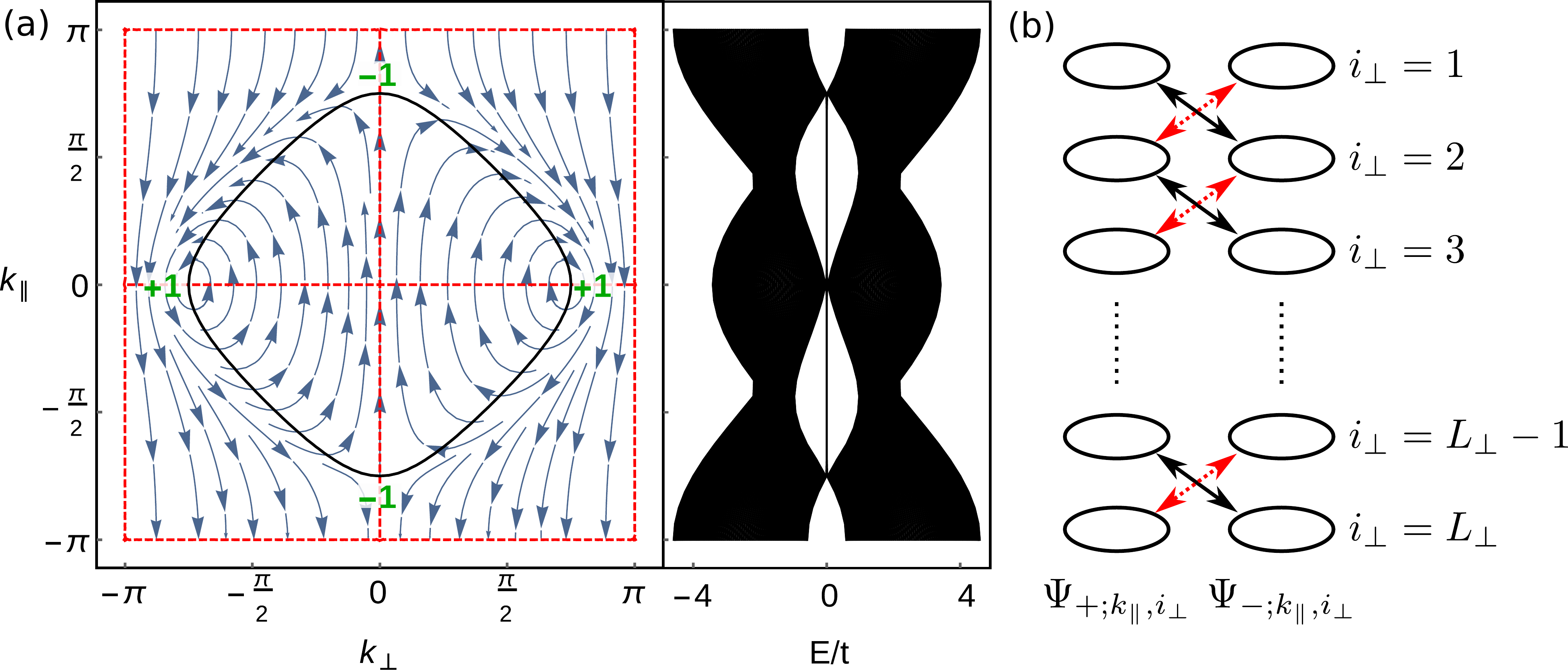}
\caption{\label{FreeSpec}  
(Color online) 
(a) Wavefunction topology of \eq{Bulk_model_SM}. The left part shows the normal state Fermi surface (black, solid), the nodal lines of $\Delta_{\bf k}$ (red, dashed), the phase of $\varepsilon_{\bf k} + i \Delta_{\bf k}$ (blue arrows), and the topological charge of the bulk nodes (green). 
The right part shows the edge spectrum containing zero-energy flat-bands.
(b) Visualization of \eq{Bulk_model_chiral}: Hopping along the black (dashed red) bonds for $s_{k_\parallel}=\sgn(t\Delta_{k_\parallel})$ positive (negative); unpaired zero-energy modes $ \Psi_{\pm;k_\parallel,i_\perp=1}^{\dagger}$ and $ \Psi_{\mp;k_\parallel,i_\perp= L_{\perp} }^{\dagger }$
at the ends of the chain.
}  
\end{figure}

To uncover the topological properties of the nodal $d_{xy}$-wave SC given in terms of
the Bogoliubov-de Gennes Hamiltonian $\mathcal{H}_0 = \sum_{\bf k}\Psi_{\bf    k}^{\dagger}H^{\ }({\bf k})\Psi^{\ }_{\bf k}$,
with the Nambu spinor $\Psi_{\bf k} = ( c_{{\bf k} \uparrow},  c^\dag_{-{\bf k} \downarrow}  )^{\mathrm{T}}$ and
\begin{eqnarray}
\label{Bulk_model_SM}
H({\bf k}) = \begin{pmatrix}
\varepsilon_{\bf k} &
\Delta_{\bf k} \cr
\Delta^{\ast}_{\bf k} &  
- \varepsilon_{-\bf k} 
\end{pmatrix} ,
\end{eqnarray}
we decompose this two-dimensional system into a set of one-dimensional chains.
The one-dimensional subsystems are indexed by $k_\parallel$ and described by the Hamiltonian  $\mathcal{H}_{k_\parallel} = \sum_{k_\perp}\Psi_{\bf    k}^{\dagger}H^{\ }_{k_\parallel}(k_\perp)\Psi^{\ }_{\bf k}$ with
\beq
\label{1D_Bulk_SM}
H_{k_\parallel}(k_\perp) = 
-(2t\cos (k_\perp )+\mu_{k_\parallel})\tau_z +\Delta_{k_\parallel}\sin (k_\perp ) \tau_x\,.
\eeq
Within each chain, there exist two anti-unitary symmetries, a commuting TRS $\mathcal{T}_{k_\parallel} = U_T \mathcal{K}$ and an anti-commuting PHS $\mathcal{C}_{k_\parallel} = U_C \mathcal{K}$.
$\mathcal{K}$ refers to the complex conjugation which inverts only $k_\perp$ ($\mathcal{K}\Psi_{k_\parallel,k_\perp}\mathcal{K}=\Psi_{k_\parallel,-k_\perp}$).
The anti-unitary symmetries act on the 1D Hamiltonian $H_{k_\parallel}(k_\perp)$ as $U_{T,C}^\dagger H_{k_\parallel}(k_\perp) U_{T,C} = \pm H^\ast_{k_\parallel}(-k_\perp)$, where $U_T = -\tau_z$ and $U_C=i\tau_x$.
Both $\mathcal{C}_{k_\parallel}$ and $\mathcal{T}_{k_\parallel}$ square to $+\mathbf{1}$\cite{footnote1Dxy},
hence each chain falls into class BDI that can exhibit non-trivial topology in one dimension\cite{Schnyder2008,Matsuura2013}.
In fact, $H_{k_\parallel}(k_\perp)$ represents a Kitaev chain with  $\mu_{k_\parallel}=\mu+2t\cos(k_\parallel)$ and $\Delta_{k_\parallel}=\Delta_{d_{xy}}\sin(k_\parallel)$\cite{Kitaev2001}. 
This system is topologically non-trivial if $2\left|t\right|>\left|\mu_{k_\parallel}\right|$ and $\Delta_{k_\parallel}\neq0$. Its topology is classified by $W_{k_\parallel}=(2\pi i )^{-1}\int_0^{2\pi}d\,k_{\perp} \partial_{k_\perp}\ln(q_{\bf k})$, which measures the winding of the phase of $q_{\bf k}=\varepsilon_{\bf k} + i \Delta_{\bf k}$ [see \fig{FreeSpec}(a)]\cite{Schnyder2012}. 

Before we derive the analytical form of the topological protected zero-energy bound states, we present a heuristic argument for their existence.

Here, it is useful to distinguish weak-pairing ($2\left|t\right|>\left|\mu_{k_\parallel}\right|$) and strong-pairing ($2\left|t\right|<\left|\mu_{k_\parallel}\right|$). For the later, we can adiabatically connect the SC state to the normal state with $\Delta_{k_\parallel}=0$, that actually is a band insulator. Hence the strong-paring case is topologically trivial.

In the weak-pairing situation, we can adiabatically tune the parameters to the high-symmetry point  $(\mu_{k_\parallel} , \Delta_{k_\parallel} ) =   (0, 2t \sgn[t\Delta_{k_\parallel}] )$. Since $q_{k_\parallel}=-2t\exp (-i \sgn[t\Delta_{k_\parallel}] k_\perp )$ winds once around the origin the chain has non-trivial bulk topology, provided that $\Delta_{k_\parallel}~\neq~0$, see Fig.~1(a).

We visualize the Majorana edge states by Fourier transforming  $H_{k_\parallel}$ with respect to $k_{\perp}$ and obtain
\beq
\label{Bulk_model_chiral}
\mathcal{H}_{k_\parallel} \sim -2 t \sum_{i_\perp} \Psi_{+s_{k_\parallel};k_\parallel,i_\perp}^{\dagger}\Psi_{-s_{k_\parallel};k_\parallel,i_\perp +1}^{\phantom{\dagger}} +h.c.\,,
\eeq
with the short hand notation $s_{k_\parallel}=\sgn(t\Delta_{k_\parallel})$ and the
chiral Majorana operators $\Psi^\dagger_{\pm;k_\parallel,i_\perp}=\frac{1}{\sqrt{2}}(c^\dagger_{k_\parallel,i_\perp \uparrow} \pm i  c^{\phantom{\dag}}_{-k_\parallel,i_\perp \downarrow})$.
The Majorana operators $\Psi_{\pm;k_\parallel}$ are eigenoperators of the chiral symmetry $\mathcal{S}_{k_\parallel} =\mathcal{C}_{k_\parallel} \mathcal{T}_{k_\parallel} =-\tau_y$ with eigenvalue $\pm 1$. Hamiltonian~\eqref{Bulk_model_chiral} consists of a chain of decoupled pairs of Majorana operators with opposite chiral eigenvalue, as  illustrated  in \fig{FreeSpec}(b).
For open boundary conditions, the Majorana operators $ \Psi_{-s_{k_\parallel};k_\parallel,i_\perp=1}^{\dagger}$ and $ \Psi_{+s_{k_\parallel};k_\parallel,i_\perp= L_{\perp} }^{\ }$ are unpaired, realizing Majorana zero modes localized at the ends of the chain.

Tuning the parameters $(\mu_{k_\parallel} , \Delta_{k_\parallel} )$ away from the high symmetry point, the edge modes acquire a finite decay length and are now described by $\gamma^\dagger_{k_\parallel} = \sum_{i_\perp} \phi_{k_\parallel}(i_\perp) \Psi_{-s_{k_\parallel};k_\parallel,i_\perp}^{\dagger}$, with the wave function $\phi_{k_\parallel}(i_\perp)$\cite{Kitaev2001}. This result is derived in the remaining part of this section

The symmetry $\mathcal{S}_{k_\parallel}$ allows to classify zero energy edge states by their chirality and we therefore use the chiral basis $\left| \pm \right\rangle=\frac{1}{\sqrt{2}}(1,\mp i)^T$ with $\mathcal{S}_{k_\parallel}\left| s \right\rangle=s\left| s \right\rangle$. This leads to the ansatz $\Phi_s(y) = e^{\kappa_s \, y} \left| s \right\rangle$.
The equation $\mathcal{H}_{k_\parallel}\Phi_s(y) = E \Phi_s(y)$ for $E=0$ generates the secular equation
\begin{subequations}
\beq
0 = 2 \frac{t_{k_\parallel}}{\Delta_{k_\parallel}} \cosh (\kappa_s) +\frac{\mu_{k_\parallel}}{\Delta_{k_\parallel}} + s \sinh (\kappa_s) ,
\eeq
that determines $\kappa_{s,\alpha}$
\beq
e^{\kappa_{s,\pm}}=\frac{-\mu_{k_\parallel} \pm \sqrt{\Delta_{k_\parallel}^2  - (4 t^2 - \mu_{k_\parallel}^2)}}{2t + s \Delta_{k_\parallel}}\,.
\eeq
\end{subequations}

To fulfil the boundary conditions $\Phi_s(y=0)=0$ and $\Phi_s(y\rightarrow\infty)=0$ for a half-infinite geometry, the wave function has to be proportional to $e^{\kappa_{s,+}y} - e^{\kappa_{s,-}y}$. Additionally, normalizability requires that both $\left|e^{\kappa_{s,\pm}}\right|$ are either smaller or larger than $1$. The former (latter) is then localized around $y=1$ ($y=L_\perp$).
In the weak paring limit, we can use $\left| \sqrt{ \Delta_{k_\parallel}^2  - (4 t^2 - \mu_{k_\parallel}^2)} \right| < \left|\Delta_{k_\parallel}\right| $ to approximate $\left| e^{\kappa_{s,\pm}} \right|  < \frac{ \left| 2 t \vphantom{\Delta_{k_\parallel}} \right| + \left| \Delta_{k_\parallel} \right| }{ \left| 2t + s \Delta_{k_\parallel} \right|}$. Hence, the chirality $s=\sgn \left( t \Delta_{k_\parallel} \right)$ state is exponentially localized around $y=1$, whereas the state of opposite chirality is localized on the other edge, which can be inferred from the relation $e^{\kappa_{+,\pm}}=e^{-\kappa_{-,\mp}}$. 

From now on, we focus on the top edge ($y=1$) and introduce the creation operator $\gamma_{k_\parallel}^\dagger$ for the according bound state $\Phi_{k_\parallel}$ with momentum $k_\parallel$ and chirality $s_{k_\parallel}=\sgn \left( t \Delta_{k_\parallel} \right)$
\begin{subequations}
\label{SurStateAnalytic}
\beqarray
\label{SurStateAnalyticOp}
\gamma^\dagger_{k_\parallel} &=&\sum_{i_\perp=1}^{L_\perp} \phi_{k_\parallel} (i_\perp) \frac{1}{\sqrt{2}}(c^\dagger_{k_\parallel,i_\perp \uparrow} - i \, s_{k_\parallel}\, c^{\phantom{\dag}}_{-k_\parallel,i_\perp \downarrow})  , \\
\phi_{k_\parallel}(y) & = & \mathcal{N}^{-1}(e^{y\kappa_{s_{k_\parallel},+}} - e^{y\kappa_{s_{k_\parallel},-}}) \,  ,
\eeqarray
\end{subequations}
with the normalization $\mathcal{N}^2=\sum_{y=1}^{L_\perp} \left| e^{y\kappa_{s_{k_\parallel},+}} - e^{y\kappa_{s_{k_\parallel},-}}\right|$.
As $e^{\kappa_{s_{k_\parallel},\pm}}$ are either both real or a complex conjugate pair, $\phi_{k_\parallel}(y)$ can be chosen to be real, which is assumed from now on. We also observe that $\phi_{-k_\parallel}(y)=\phi_{k_\parallel}(y)$.

These edge states are charge neutral, carry a spin of $S_z=+1$ and their chirality is locked to the momentum as $\sgn(k)$. In analogy to the edge states of a quantum-spin-hall system, the state with opposite chirality is bound to the second edge at infinity. Observe that the neutral edge states can still carry an electrical current as the electron-like contribution propagates in the opposite way as the hole-like part. In contrast, it cannot contribute to spin currents along the edge.

The flatness of the Majorana fermions ($E=0$) is protected by the standard TRS ($c_{\bf k} \rightarrow i\sigma_y c_{-\bf k}$) and translation symmetry along the edge. All possible mass terms are given by
\beq
\label{EdgeSplittingGeneral}
\mathcal{H}_q = \sum_{k_\parallel} \left[ a_q(k_\parallel) \gamma_{k_\parallel}^{\dagger}\gamma_{k_\parallel + q}^{\ } + b_q(k_\parallel) \gamma_{-k_\parallel}^{\ } \gamma_{k_\parallel + q}^{\ } +h.c. \right] .
\eeq
The edge state operators transform under TRS as $\gamma_{k_\parallel}^{\dagger}\rightarrow-is_{k_\parallel}\gamma_{k_\parallel}^{\ }$ and \eq{EdgeSplittingGeneral} accordingly as
\beqarray
\label{EdgeSplittingTRS}
&& \mathcal{H}_q  \rightarrow - \sum_{k_\parallel} \sgn\left( \sin(k_\parallel)\sin(k_\parallel+q)\right)  \\
& & \qquad
\times \left[ a_q(k_\parallel) \gamma_{k_\parallel}^{\dagger}\gamma_{k_\parallel + q}^{\ } + b_q(k_\parallel) \gamma_{-k_\parallel}^{\ } \gamma_{k_\parallel + q}^{\ } +h.c. \right] . \nonumber
\eeqarray
All homogeneous mass terms with $q=0$ break only TRS, whereas all other terms with $q \ne 0, \pi$ break both TRS and translation symmetry. The instability with $q=\pi$ is special, since it only breaks translation, but not TRS.

\section*{II.~~~Projection onto Edge States}
\label{Projection}

Here, we decompose the fermion operators $\Psi_{k_\parallel,i_\perp}=( c_{k_\parallel,i_\perp; \uparrow},  c^\dag_{-k_\parallel,i_\perp; \downarrow}  )$ in terms of the eigenstates $\eta_{k_\parallel,n}$ of the non-interacting system
\begin{subequations}
\beqarray
\eta_{k_\parallel,n}&=&\sum_{i_\perp,\tau}U_{n,(i\perp,\tau)}(k_\parallel)\Psi_{k_\parallel,i_\perp;\tau}\\
\Psi_{k_\parallel,i_\perp;\tau}&=&\sum_{n}U^\dagger_{(i\perp,\tau),n}(k_\parallel)\eta_{k_\parallel,n}\,.
\eeqarray
\end{subequations}
To project onto the edge states, we only keep the $n=0$ contributions, with $\eta_{k_\parallel,0}=\gamma_{k_\parallel}$, and ignore all other parts:
\begin{subequations}
\label{Projections}
\beqarray
c^{\ }_{k_\parallel,i_\perp,\uparrow} 
& \rightarrow & \frac{1}{\sqrt{2}} \phi_{k_\parallel}(i_\perp)\gamma^{\ }_{k_\parallel} 
\\
c^{\ }_{k_\parallel,i_\perp,\downarrow} 
& \rightarrow & \frac{-i\,s_{k_\parallel}}{\sqrt{2}} \phi_{k_\parallel}(i_\perp)\gamma^\dagger_{-k_\parallel} 
\eeqarray 
\end{subequations}
By substituting \eq{Projections} into the definition of the physical spin operator, we obtain the  projected versions
\beq
\label{ProjPhysSpin}
{\bf{S}}_0(i_\perp) =  \sum_{k_\parallel=0}^\pi \phi_{k_\parallel}^2(i_\perp) \Gamma_{k_\parallel}^\dagger \frac{\btau}{2} \Gamma_{k_\parallel}^{\ }\, ,
\eeq
where we have introduced the basis $\Gamma_{k_\parallel}^\dagger=(\gamma_{k_\parallel}^\dagger,-i\,s_{k_\parallel}\,\gamma_{-k_\parallel}^{\ })$.

Substituting \eq{SurStateAnalyticOp} into \eq{Projections} nicely demonstrates the consequences of the projection onto chiral edge states through the replacement rules:
\begin{subequations}
\label{ProjectionReversed}
\beqarray
\hspace{-0.5cm}
c^{\ }_{k_\parallel,i_\perp,\uparrow} 
& \rightarrow & \frac{\phi^2_{k_\parallel}(i_\perp)}{2} \left( c^{\ }_{k_\parallel,i_\perp,\uparrow} + i s_{k_\parallel} c^{\dagger }_{-k_\parallel,i_\perp,\downarrow} \right) + \dots
\\
\hspace{-0.5cm}
c^{\ }_{k_\parallel,i_\perp,\downarrow} 
& \rightarrow & \frac{\phi^2_{k_\parallel}(i_\perp)}{2}  \left( c^{\ }_{k_\parallel,i_\perp,\downarrow} - i s_{k_\parallel} c^{\dagger }_{-k_\parallel,i_\perp,\uparrow} \right)+ \dots
\eeqarray 
\end{subequations}
In the above, we kept only contributions at the original position $i_\perp$. Additional terms due to the sum in \eq{SurStateAnalyticOp} are represented by $(\dots)$. This analysis demonstrates the level at which normal and SC order are intertwined. If the edge supports another state with the same wave function $\phi_{k_\parallel}(i_\perp)$ of opposite chirality, the anomalous contribution $c^\dagger$ cancels and the only consequence of the projection is a prefactor of $\phi^2_{k_\parallel}(i_\perp)$. Hence, the SC ground state may also (dynamically) mix normal and SC order parameter, but this mixing takes place on a different level.

Expanding $\phi^4_{k_\parallel}(i_\perp)/2=a_0 + \dots$ and  $s_{k_\parallel}\phi^4_{k_\parallel}(i_\perp)/2=2b_1 \sin(k_{k_\parallel}) + \dots$ in harmonic functions and using the above relations, we find the following  decompositions
\begin{subequations}
\beqarray
S_0^{x\vphantom{(\Psi)}}
& =
& \sum_j \left[ a_0 S^x_j + b_1(\Delta_j^{b,x} + {\Delta_j^{b,x}}^\dagger) \right] + \dots \\ 
S_0^{y\vphantom{(\Psi)}} 
& = 
& \sum_j \left[ a_0 S^y_j + b_1(\Delta_j^{b,y} + {\Delta_j^{b,y}}^\dagger) \right] + \dots \\ 
S_0^{z\vphantom{(\Psi)}} 
& =
& \sum_j \left[ a_0 S^z_j - b_1(\Delta_j^{b,z} + {\Delta_j^{b,z}}^\dagger) \right] + \dots
\eeqarray
\end{subequations}

The above derivation assumed half filling, such that the SC nodes are located in the edge Brioullin zone at $0$ and $\pi$. The analysis itself however does not crucially depend on this assumption. Doping the system away from half filling shortens the flat band and the summation in \eq{ProjPhysSpin} has to be adapted accordingly. Nevertheless, the edge states still come in pairs $(k_\parallel,-k_\parallel)$ and there is again a mixing of normal and SC operators. The only point that requires more work is the decomposition in harmonic functions and the Fourier transformation that lead to the equations above.

In the following calculations, we use the enhanced symmetry at half filling explicitly. Here, the sublattice symmetry guaranties the relation $\phi_{k_\parallel}(i_\perp)=-(-1)^{i_\perp}\phi_{k_\parallel+\pi}(i_\perp)$. As the interaction is restrained to $i_\perp=1$ and the QMC study is performed in this layer only, we drop the sign completely.
The projection of ${\bf S}^{(\Psi)}_q$ onto the edge states vanishes for the $x$- and $z$-component with $q=0$ and for the $y$-component with $q=\pi$. The three non-vanishing parts $(S_\pi^{x,(\Psi) }, S_0^{y,(\Psi) }, S_\pi^{z,(\Psi) })^T$ generate the Majorana mass terms with $\tilde{\Gamma}_{k_\parallel}^\dagger=(\gamma_{k_\parallel}^\dagger,-i\,s_{k_\parallel}\,\gamma_{k_\parallel-\pi}^{\dagger })$
\beq
(S_\pi^{x,(\Psi) }, S_0^{y,(\Psi) }, S_\pi^{z,(\Psi) })^T =  \sum_{k_\parallel=0}^\pi \phi_{k_\parallel}^2(i_\perp) \tilde{\Gamma}_{k_\parallel}^\dagger \frac{\btau}{2} \tilde{\Gamma}_{k_\parallel}^{\ }\, .
\eeq
We obtain the projected fermionic operator by substituting \eq{ProjectionReversed} into the definition of ${\bf S}^{(\Psi)}_q$:
\begin{subequations}
\beqarray
S_\pi^{x(\Psi)}  
& =
& \sum_j (-1)^j \left[ a_0 (\Delta_j^{s} + {\Delta_j^{s}}^\dagger)  + b_1 n^{b}_j\right]  + \dots \\ 
S_0^{y(\Psi)} 
& =
& \sum_j  \left[ - i a_0 (\Delta_j^{s} - {\Delta_j^{s}}^\dagger) + b_1 J_j \right] + \dots \\ 
S_\pi^{z(\Psi)} 
& =
& \sum_j (-1)^j \left[ a_0 n_j - b_1(\Delta_j^{b,s} + {\Delta_j^{b,s}}^\dagger) \right] + \dots \quad \quad
\eeqarray
\end{subequations}

\section*{III.~~~QMC Method}

For the numerical simulations we employ the action based continuous-time Quantum-Monte-Carlo method in the interaction expansion\cite{Rubtsov2005,Gull2011} which stochastically samples the grand canonical partition function $Z$ using a Metropolis-Hastings algorithm.
To start we introduce the Gaussian part $S_0$ and the interacting part $S_{I}$ of the action $S$ as
\beqarray
S_0 &=& -\sum_{{\bf i},{\bf j}}\iint_0^\beta d\tau \, d\tau ' \Psi^\dagger_{{\bf i},\tau} G_0^{-1}({\bf i - j}, \tau - \tau') \Psi_{{\bf j},\tau'}\quad\quad \\
S_{I} &=& - U \sum_{i_e}\int_0^\beta \prod_\sigma (\Psi^\dagger_{\sigma,i_e,\tau}\Psi_{\sigma,i_e,\tau}-\frac{1}{2})\, ,
\eeqarray
where $G_0^{-1}({\bf i - j}, \tau - \tau')$ is the free BdG-Greens function of the two-dimensional system \eq{Bulk_model_SM} in ribbon geometry.

\begin{figure}
\begin{minipage}{.5\columnwidth}
\centering
\includegraphics[clip,angle=0,width=\columnwidth]{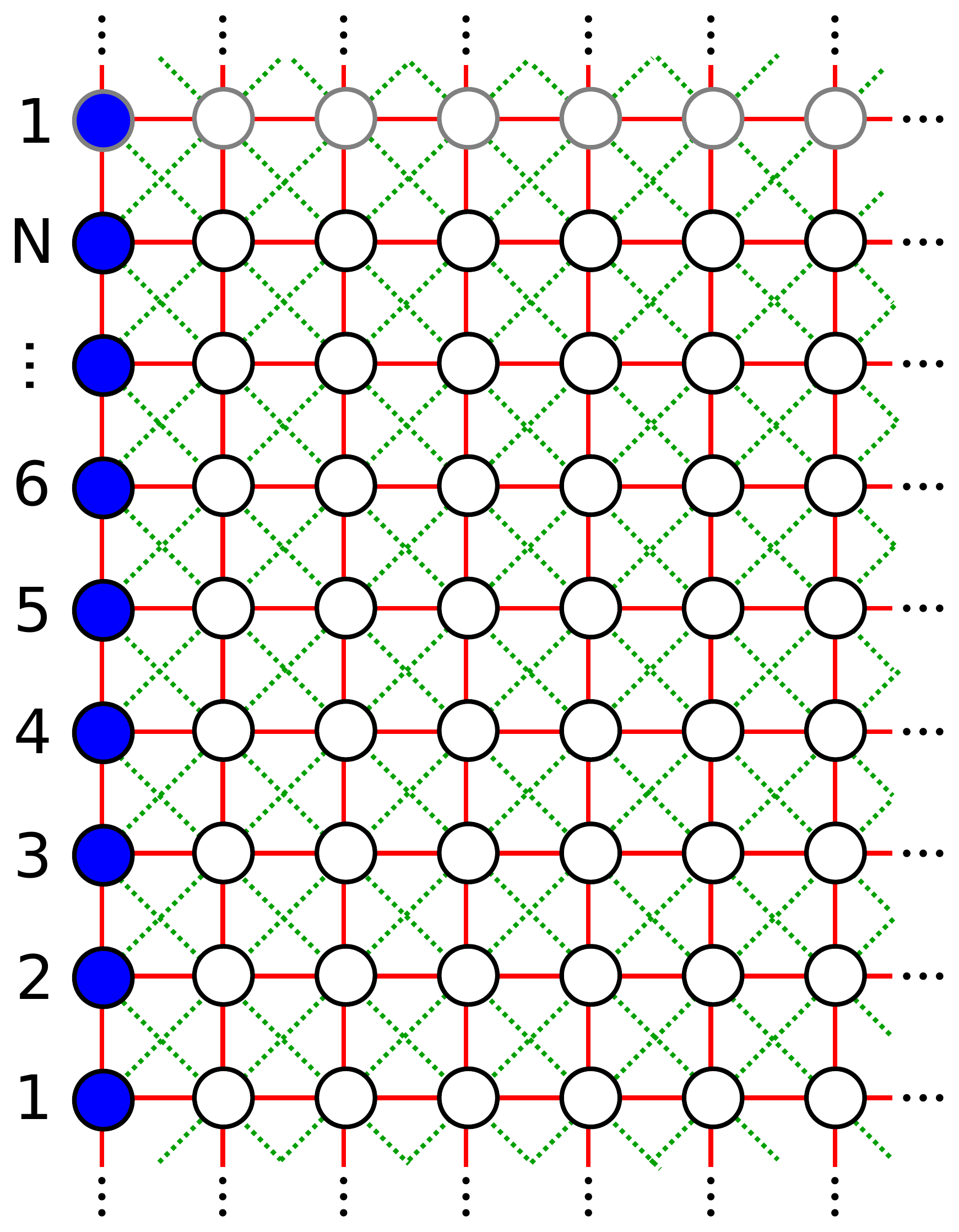}
\end{minipage}
\hfill
\begin{minipage}{.45\columnwidth}
\caption{\label{Lattice}  
(Color online) 
Visualization of the square lattice in ribbon geometry: normal hopping terms along nearest neighbour bond (solid red), superconducting $d_{xy}$-wave pairing on next-nearest neighbour bonds (dashed greed), on-site chemical potential in the bulk and interactions along the edge (filled blue circles). We assume periodic boundary conditions in the parallel direction and open once for the perpendicular direction
}  
\end{minipage}
\end{figure}

To proceed we introduce the grand canonical partition function $Z$ in terms of the action as
\beqarray
Z & = &\Tr\left[ e^{-\beta (\mathcal{H}_0+\mathcal{H}_{int})} \right] \\
  & = & Z_0 \sum_n \frac{\left( -1 \right )^n}{n!}  \left \langle  
S_{I}{}^n \right \rangle_0 \, ,
\eeqarray
where we have used the definition of time-ordered expectation value $\left \langle \dots \right \rangle_0 = Z_0^{-1} \int \mathcal{D}[\Psi^\dagger,\Psi] \left[ T_\tau \dots e^{-S_0} \right]$ with $Z_0=\Tr\left[ e^{-\beta \mathcal{H}_0} \right]$ being the partition function of the non-interaction system.

For the following discussion it is useful to define two shorthand notations; firstly $v_j$ for the $j$-th vertex $v_j=\prod_\sigma (\Psi^\dagger_{\sigma,i_{e_j},\tau_j}\Psi_{\sigma,i_{e_j},\tau_j}-\frac{1}{2})$ at position $(i_{e_j},\tau_j)$ and secondly the superindex $C_n$ for a configuration of order $n$ containing all internal positions of the vertices $C_n=\lbrace(i_{e_1},\tau_1),\dots,(i_{e_n},\tau_n)\rbrace$.
Hence, the partions function is given as
\beq
\frac{Z}{Z_0} = \sum_{C_n} \frac{U^n}{n!} \left \langle v_1 \dots v_n \right \rangle_0\,.
\eeq

The expectation value $\left \langle \dots \right \rangle_0$ is taken with respect to the non-interacting theory, hence we can use Wicks theorem within each individual configuration to contract the vertices.
This nicely visualizes the QMC algorithm at hand as a random walk through the space of all possible Feynman diagrams. For the Metropolis-Hastings updates, we either propose to add a vertex at a randomly chose position or to remove one arbitrary vertex of the configuration which stochastically samples the partition function without any cutoffs, for example in the expansion order.

As the interaction is restricted to the edge, the evaluation of $\left \langle v_1 \dots v_n \right \rangle_0$ will exclusively evoke propagators between two edge sites. Accordingly, the simulation appears to be one-dimensional. Nevertheless the Greens function still contains the information about all possible paths in the original two-dimension system and thereby respects all degrees of freedom including bulk states.


\begin{thebibliography}{54}
\bibitem [{foo({\natexlab{a}})}]{footnoteA}%
  \BibitemOpen
  \href@noop {} {} \bibinfo {note} {Note that splitting of
  the zero-bias peak at low temperatures was not confirmed in all experiments,
  see, e.g., \onlinecite{kashiwaya_prb_04}.}\BibitemShut {Stop}%
\end{thebibliography}

\end{document}